\documentclass[10pt]{article}

\usepackage{geometry}
\usepackage{amsthm,amsmath}
\usepackage[square]{natbib}
\setcitestyle{numbers,comma}
\usepackage{latexsym, amsfonts, amssymb, amscd}
\usepackage{bbm}
\usepackage[T1]{fontenc}
\usepackage{graphicx}
\usepackage[utf8]{inputenc}
\usepackage{abstract}

\usepackage{titlesec}

\titlespacing*{\section}
  {0pt}
  {0.5ex plus .2ex minus .2ex}
  {0.5ex plus .2ex}
  [0pt]

\makeatletter
\renewenvironment{thebibliography}[1]
     {\section*{\refname}%
      \@mkboth{\MakeUppercase\refname}{\MakeUppercase\refname}%
      \list{\@biblabel{\@arabic\c@enumiv}}%
           {\settowidth\labelwidth{\@biblabel{#1}}%
            \leftmargin\labelwidth
            \advance\leftmargin\labelsep
            \@openbib@code
            \usecounter{enumiv}%
            \let\p@enumiv\@empty
            \renewcommand\theenumiv{\@arabic\c@enumiv}}%
      \setlength{\itemsep}{0cm}
      \sloppy
      \clubpenalty4000
      \@clubpenalty \clubpenalty
      \widowpenalty4000%
      \sfcode`\.\@m}
     {\def\@noitemerr
       {\@latex@warning{Empty `thebibliography' environment}}%
      \endlist}
\makeatother

\usepackage{titling}
\pretitle{\begin{center}\LARGE}
\posttitle{\par\end{center}\vspace{0.5em}}
\preauthor{\begin{center}\small
           \begin{tabular}[t]{c}}
\postauthor{\end{tabular}\par\end{center}}

\newcommand{\expect}[1]{\left< #1 \right>}

\newcommand{\Isyn}{I^\mathrm{syn}}
\newcommand{\taueff}{\tau^\mathrm{eff}}
\newcommand{\taum}{\tau^\mathrm{m}}
\newcommand{\taubk}{\tau^\mathrm{b}_k}
\newcommand{\tauref}{\tau^\mathrm{ref}}
\newcommand{\tausyn}{\tau^\mathrm{syn}}

\title{
The high-conductance state enables\\neural sampling in networks of LIF neurons
}

\date{26.02.2015}
\author{Mihai A. Petrovici, Ilja Bytschok, Johannes Bill, Johannes Schemmel and Karlheinz Meier\thanks{
This research was supported by EU grants \#269921 (BrainScaleS), \#237955 (FACETS-ITN), \#604102 (Human Brain Project), the Austrian Science Fund FWF \#I753-N23 (PNEUMA) and the Manfred St\"ark Foundation.}
}

\begin{document}

\maketitle

The apparent stochasticity of \textit{in-vivo} neural circuits has long been hypothesized to represent a signature of ongoing stochastic inference in the brain \citep{kording2004bayesian, fiser2010statistically, friston2011action}.
More recently, a theoretical framework for \textit{neural sampling} has been proposed, which explains how sample-based inference can be performed by networks of spiking neurons \citep{buesing2011neural,petrovici2013stochastic}.
One particular requirement of this approach is that the membrane potential of these neurons satisfies the so-called \textit{neural computability condition (NCC)}, which in turn leads to a \textit{logistic neural response function} $\nu_k(\bar u_k)$:
\begin{equation}
    \bar u_k = \ln \frac{p(z_k=1|z_{\backslash k})}{p(z_k=0|z_{\backslash k})} \quad \Longrightarrow \quad \nu_k \propto p(z_k=1|z_{\backslash k}) = \frac{1}{1+\exp(-\bar u_k)} \quad ,
\end{equation}
where $\bar u_k$ represents the mean free membrane potential of the neuron.

Analytical approaches to calculating neural response functions have been the subject of many theoretical studies.
In order to make the problem tractable, particular assumptions regarding the neural or synaptic parameters are usually made.

\begin{figure}[t]
    \centering
    \includegraphics[width=\columnwidth]{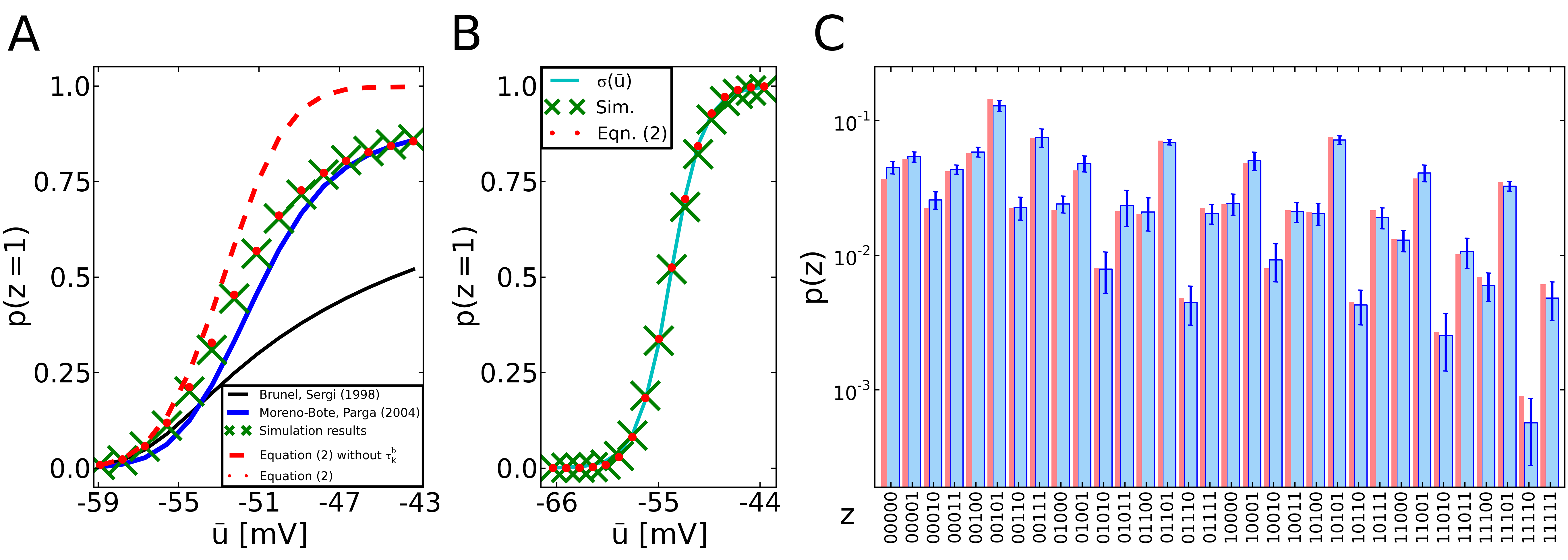}
    \caption{
        (A) Prediction of the neural response function for an LIF neuron under Poisson bombardment of intermediate strength.
        Existing theories (black and blue curves) do not hold in this regime, due to particular assumed approximations.
        Our theory, based on a propagation of the membrane autocorrelation throughout spike bursts (red dots), accurately reproduces simulation results (green crosses).
        The theoretical approach presented here represents an extension of the one we previously developed in \cite{petrovici2013stochastic} (dashed red curve).
        (B) Prediction of the neural response function for an LIF neuron in the HCS (red dots) and simulation data (green crosses).
        In this regime, the response function becomes a (linearly transformed) logistic function (cyan curve).
        (C) A recurrent network of 5 LIF neurons in the HCS sampling from a (randomly drawn) Boltzmann distribution.
        The joint probability distribution sampled by the network after $10^4$ ms (blue bars) is compared to the target distribution (red bars).
        Errors are calculated over 10 different trials.
        \label{fig:fig}
    }
\end{figure}

One common assumption is that synaptic time constants are negligible ($\tausyn \to 0$) compared to the membrane time constant, which, in the diffusion approximation, allows treating the membrane potential as an \textit{Ornstein-Uhlenbeck (OU) process} \cite{ricciardi1979ouprocess}.
The particular appeal of this formal equivalence is that it allows calculating the mean first-passage time $\expect{T(\vartheta,\varrho)}$ of the neural membrane potential $u$ from the reset potential $\varrho$ to the threshold potential $\vartheta$ \cite{ricciardi1988fptdensity}, which is equivalent to the inverse of the firing rate $\nu$.
This approach can be further refined by considering nonzero, but small $\tausyn$ by means of an expansion in $\sqrt{\tausyn/\taum}$ \cite{brunel1998firingfrequency}.
However, biologically significant activity regimes exist which are not covered by the underlying assumption:
Under strong synaptic bombardment, as is often the case in cortex, the summation of synaptic conductances shifts the neuron into a \textit{high-conductance state (HCS)}, which is characterized by a very fast effective membrane time constant $\taueff$.
In this case, the opposite limit may appear ($\sqrt{\tausyn/\taueff} \to \infty$), and the correlation between the membrane potential before and after an output spike becomes significant (Figure \ref{fig:fig}A, black curve).

The specific limit of $\tausyn \gg \taum$ has also been studied \cite{moreno2004role}, albeit with a different approach.
In this limit, an \textit{adiabatic (or quasistatic) approximation} can be made, since the membrane reacts much faster than the synaptic input changes.
This allows calculating the response function as an integral of the firing rate $\tilde\nu(I)$ for constant input current $I$ over the probability density function (PDF) $p(\Isyn)$, which, for an OU process, can be given in closed form.
The adiabatic assumption, however, also represents the limitation of this approach.
When the synaptic time constants are in the order of the refractory period $\tauref$, the synaptic input current is no longer quasistatic with respect to the time of spiking compared to the time when the membrane is released from the reset potential (Figure \ref{fig:fig}A, blue curve).

The regime in which $\taueff \ll \tausyn\sim\tauref$ is, however, not only biologically relevant, but also particularly interesting from a functional point of view.
In \cite{petrovici2013stochastic}, we have shown that LIF neurons that are shifted into a HCS by background synaptic bombardment can attain the correct firing statistics (i.e., satisfy the NCC) to sample from well-defined probability distributions.
This framework builds on the abstract model from \cite{buesing2011neural}, which explicitly establishes an approximate equivalence of $\tauref$ and $\tausyn$.
In order to calculate the response function of neurons in this regime, we were therefore required to consider a different approach.
Here, we present an extended version of this theory, which remains valid for a larger parameter space.

The core idea of this approach is to separately consider two different ``modes'' of spiking dynamics: burst spiking, where the effective membrane potential is suprathreshold and the inter-spike-intervals are close to the refractory time, and transient quiescence, in which the neuron does not spike for longer periods.
For the bursting mode, we explicitly take into consideration the autocorrelation of the membrane potential before and after refractoriness by \textit{propagating the PDF} of the effective membrane potential from spike to spike within a burst.
For the membrane potential evolution between bursts, we consider an OU-like approximation.
The response function can then be given by
\begin{equation}
    \nu_k = \frac{\sum_n n P_n \tauref}{\sum_n P_n \left(n\tauref + \sum_{k=1}^{n-1} \taubk + T_n\right)} \quad ,
\end{equation}
where $P_n$, $T_n$ and $\taubk$ represent quantities that are associated with a burst length of $n$ spikes and can be calculated recursively:
\begin{align}
    P_n     & = \textstyle \left( 1 - \sum_{i=1}^{n-1} P_i \right) \int_{\vartheta}^\infty du_{n-1} p(u_{n-1} | u_{n-1} \geq \vartheta) \left[ \int_{-\infty}^{\vartheta} du_n p(u_n | u_{n-1}) \right] \; , \\
    T_n     & = \textstyle \int_{\vartheta}^\infty du_{n-1} p(u_{n-1} | u_{n-1} \geq \vartheta) \left[ \int_{-\infty}^{\vartheta} du_n p(u_n | u_n < \vartheta, u_{n-1}) \expect{T(\vartheta, u_n)} \right] \; , \\
    \taubk &= \textstyle \taueff \int_\vartheta^\infty du_k \ln \left(\frac{\varrho-u_k}{\vartheta - u_k}\right) p(u_k | u_k > \vartheta, u_{k-1}) \; .
\end{align}

In the limit of small $\taueff$, we show that the neural activation function $p(z_k=1|\Isyn)=\nu_k\tauref$ becomes symmetric and can be well approximated by a logistic function, thereby providing the correct dynamics in order to perform neural sampling (Figure \ref{fig:fig}B).
Such stochastic firing units can then be used to sample from arbitrary probability distributions over binary random variables (RVs) $z_k$ \cite{petrovici2013stochastic,buesing2011neural,pecevski11probabilistic,probst2015probabilistic}.
Figure \ref{fig:fig}C shows a network of 5 neurons sampling from an exemplary Boltzmann distribution.
Inference in these spaces is readily performed by injecting a strong current into the neurons corresponding to observed RVs, allowing the network to sample from the resulting posterior distribution.

This spike-based sampling approach offers several important advantages: it allows an increasingly accurate representation of the underlying probability distribution at any time (``anytime computing''), marginalization comes at no cost, as it can be done by simply neglecting the values of the ``uninteresting'' or unobserved RVs, and the physical implementation of the sampling algorithm, i.e., the network structure, is massively parallel by construction.
We thereby provide not only a normative framework for Bayesian inference in cortex, but also powerful applications of low-power, accelerated neuromorphic systems to highly relevant machine learning problems.

\vspace{20pt}

\bibliographystyle{alpha}

\vspace{7pt}

\end{document}